# The SLAC Polarized Electron Source[*]

J. E. Clendenin, A. Brachmann, T. Galetto, D.-A. Luh, T. Maruyama, J. Sodja, and J. L. Turner

*Stanford Linear Accelerator Center, 2575 Sand Hill Rd., Menlo Park, CA 94025*

## Abstract

The SLAC PES, developed in the early 1990s for the SLC, has been in continuous use since 1992, during which time it has undergone numerous upgrades. The upgrades include improved cathodes with their matching laser systems, modified activation techniques and better diagnostics. The source itself and its performance with these upgrades will be described with special attention given to recent high-intensity long-pulse operation for the E-158 fixed-target parity-violating experiment.

*Presented at the Workshop on Polarized Electron Sources and Polarimeters (PESP 2002), September 4-6, 2002, Danvers, MA.*

---

[*]Work supported by Department of Energy contract DE–AC03–76SF00515.

# The SLAC Polarized Electron Source[*]

J. E. Clendenin, A. Brachmann, T. Galetto, D.-A. Luh, T. Maruyama, J. Sodja, and J. L. Turner

*Stanford Linear Accelerator Center, 2575 Sand Hill Rd., Menlo Park, CA 94025*

**Abstract**. The SLAC PES, developed in the early 1990s for the SLC, has been in continuous use since 1992, during which time it has undergone numerous upgrades. The upgrades include improved cathodes with their matching laser systems, modified activation techniques and better diagnostics. The source itself and its performance with these upgrades will be described with special attention given to recent high-intensity long-pulse operation for the E-158 fixed-target parity-violating experiment.

## 1 INTRODUCTION

The polarized electron source (PES) [1] was initially commissioned for the SLC in early 1992. The present post-SLC configuration is shown in Fig. 1. The all solid-state Nd:YLF-pumped Ti:sapphire short-pulse laser is used for filling PEP-II and for other experiments not requiring polarized electrons. The flash lamp-pumped long-pulse Ti:sapphire laser[1] has been recently improved for a fixed-target experiment[2] that requires a high-charge polarized beam. The characteristics of the PES are listed in Table 1. Which of the two laser beams is delivered to the cathode is chosen on a pulse-to-pulse basis. Improvements to the source are discussed in Section 2 and its present performance in Section 3.

## 2 IMPROVEMENTS

Given the uncertainties of operating a major accelerator with a polarized electron source, four guns were built for the following purposes: one operating, one on standby, one being repaired, and one for experiments in the off-line laboratory consisting of a duplicate of the first few meters of the 3-km linac. However, the addition of a load-lock[3] in 1993 improved the reliability of the PES to such an extent that certainly no more than 3 guns are now needed. All activations[4] are done in the load-lock with the gun isolated. Re-application of Cs to the cathode is done periodically in the gun chamber under computer control (operator initiated) with the bias voltage reduced to 1 kV. Cesium channels from SAES are used exclusively. Additional $NF_3$ is not added inbetween activations. To monitor the cathode quantum efficiency (QE), a low-power cw diode laser is mounted on a side window of the gun with a view of the cathode. The laser is modulated at a few hertz to minimize the cw beam in the linac. A fiber-optically coupled nanoammeter installed at the gun HV electrode support flange supplies a modulated current signal to a lock-in amplifier as shown in Fig. 2. The control chassis also blanks off the diode laser during the linac rf accelerating pulse to prevent any acceleration of photoelectrons generated by this source.[5]

---

[*]Work supported by Department of Energy contract DE–AC03–76SF00515.

[1]A. Brachmann *et al.*, this Workshop.
[2]P. Mastromarino *et al.*, this Workshop.
[3]Before biasing the cathode, most of the load-lock apparatus must be removed and a corona shield installed around the high-voltage (HV) electrode. Because of space constraints in the linac housing, the HV insulator and corona shield is surrounded by dry gas contained within a sealed shroud.

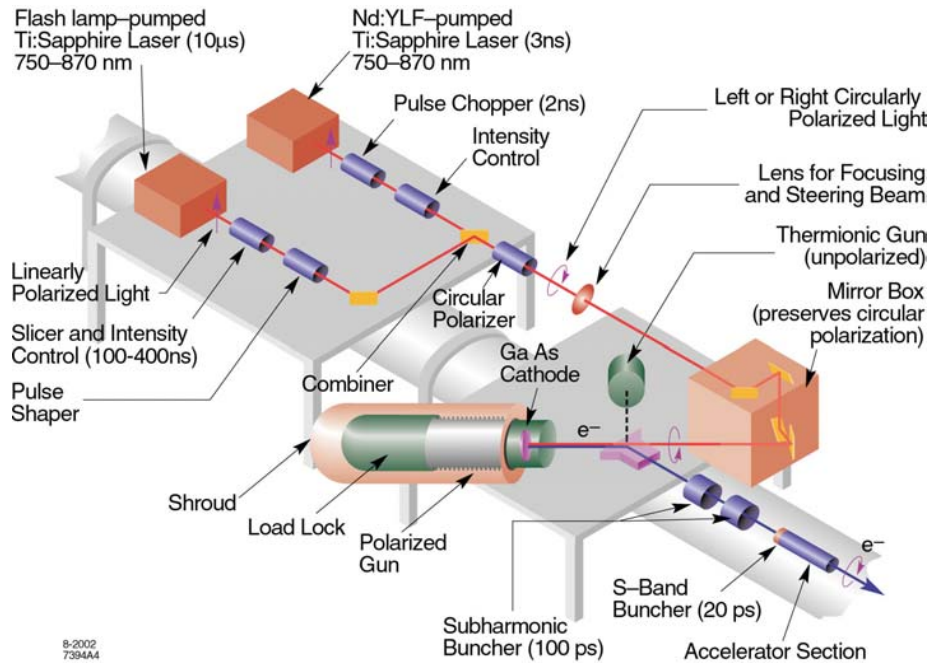

**FIGURE 1**. The SLAC polarized electron source in its present configuration.

The addition of the load-lock required the cathode to be mounted on a detachable puck that is supported by a 2.5-cm diameter emitter tube made from a 8-cm long Mo cup brazed to a Kovar tube, which is welded to a 304L stainless-steel (SST) tube. During operation, cold $N_2$ is circulated inside the tube to lower the cathode temperature to about 0 °C. When activating, a resistive heater is installed. To bring the GaAs crystal to the desired 600 °C, even with a heat shield on the vacuum side of the crystal, the puck end of the emitter tube reaches about 800°C, which formerly resulted in frequent vacuum failures of the Mo-Kovar brazed joint. Ni-plating the Kovar prior to brazing helped.[6] Eventually an ANSYS 5.1 study of the temperature response was used to determine the optimum length of the Mo cup to keep the temperature of the brazed joint below 250 °C [2]. A new Mo-length of 31 cm was chosen. Computer controlled temperature-cycling tests resulted in no failures after 70 cycles. To reduce hazards, the gas mixture in the shroud can now be readily switched from $SF_6$ to dry air or to a mixture of both. Generally $SF_6$ is used only in the first couple days of operation following closure of the shroud.

---

[4] Activations consist of heat cleaning for 1-hour at 600 °C followed by application of Cs and florine in the form of $NF_3$.
[5] The low intensity of this laser precludes a radiation issue.

**TABLE 1**. Source Characteristics

| Component | Value |
|---|---|
| Cathode diameter | 20 mm (emitting area) |
| Cathode holder | Mo puck with Ta clip |
| Electrodes | 317L SST with low carbon content, low inclusion density |
| Bias voltage alumina insulator | 39.5-cm long by 19.0-cm ID, OD fluted |
| Design/nominal bias voltage | -150/-120 kV dc |
| Dark current at nominal voltage | 30 nA typical |
| Vacuum pumping | 120 lps DI and 200 lps NEG pumps |
| Vacuum | $10^{-11}$ Torr dominated by $H_2$ |
| Ti:sapphire laser systems: | |
|   Nd:YLF pumped | 2-ns pulse for PEP II, FFTB |
|   Flashlamp pumped | Polarized 100-400 ns pulse for ESA |

The major improvement is in the cathode itself. The progress toward higher polarization is illustrated in Table 2. There has been only limited success in improving the cathode preparation technique. A vacuum transfer vessel allows the cathode to be activated in the laboratory and then transferred to the load-lock system at the gun under vacuum. The vessel has a small battery-operated IP, but nonetheless a final re-activation in the load-lock is generally necessary. A glove-box for performing the initial cathode cleaning in an $N_2$ atmosphere has also been constructed, but so far not used. Presently H* cleaning at 400 °C is being studied.[7]

# 3 PERFORMANCE

Since 1992 the polarized electron source has provided ~80% of the electron beams for the SLAC linac, accumulating nearly 40,000 h of operation with availability >>95%. Except for 1992 (when there was no load-lock) the pattern has been as follows, independent of the type of cathode used (listed in Table 2). After initial activation in the load-lock, the cathode is inserted into the gun. Neither the cathode bias HV nor the presence of the electron beam are observed to have any effect on the QE. When the QE decreases to the point where the required current can no longer be obtained—because of limited laser energy and/or because of the surface charge limit—the cathode is re-cesiated in the gun with the cathode bias reduced to –1 kV. This typically occurs every 3-5 days and takes ~20 min., the time dominated by cycling of the bias voltage. About 0.5 ML Cs is estimated to be added per re-cesiation. No $NF_3$ is added. The re-cesiation completely restores the original QE. The rate of decrease of the QE (QE lifetime) changes over long periods (generally to shorter lifetimes), but not between re-cesiations. Re-cesiations are repeated up to 50 times over the period of a year, limited only by operational considerations to preemptively change the cathode during the accelerator scheduled major mainte-nance periods. An important element in the cathode performance is the gun vacuum. An RGA scan for the present gun, which has not been back-filled for several years, is shown in Fig. 3. The RGA is operated continuously.

---

[6] Suggested by G. Collet and R. Kirby (SLAC).
[7] D.-A. Luh *et al*., this Workshop.

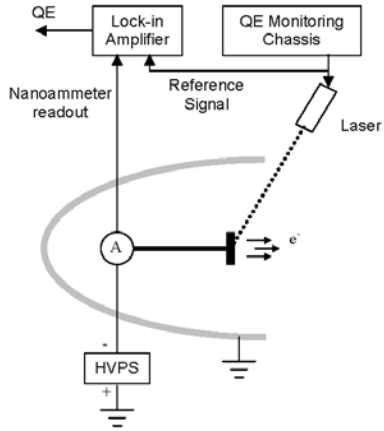

**FIGURE 2**. Use of nanoammeter (A) cathode bias for continuous monitoring of yield.

**TABLE 2**. Cathode History

| Cathode | $P_e$ (%) | Experiment |
|---|---|---|
| Bulk GaAs | 25 | 1992 SLC |
| Bulk AlGaAs | 35 | 1992 E142 |
| 300-nm strained GaAs | 65 | 1993 SLC |
| 100-nm strained GaAs | 80 | 1993 143 |
|  |  | 1994-98 SLC |
|  |  | 1995 E154 |
|  |  | 1996 E155 |
|  |  | 1999 E155X |
| Gradient-doped strained GaAs | 80 | 2001-present |

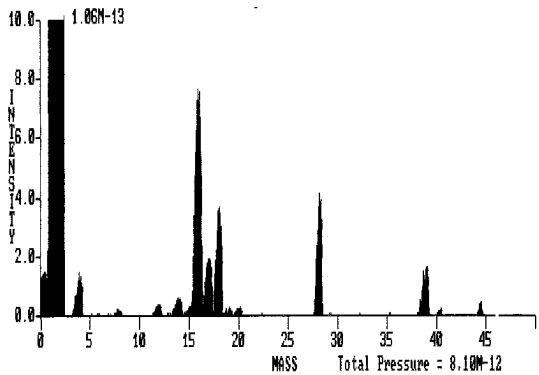

**FIGURE 3**. Recent gun-chamber RGA mass spectrum. The intensity is in units of $10^{-13}$ Torr. The m=2 (offscale) and m=4 peaks are ~$10^{-11}$ and ~$10^{-13}$ Torr respectively.

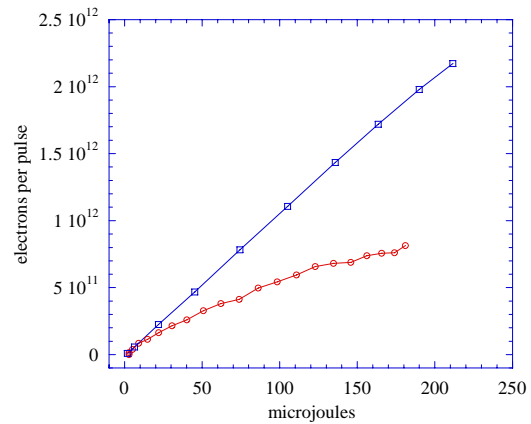

**FIGURE 4**. Comparison of emission from gradient-doped cathode (top curve, squares) using 100-ns laser pulse with medium-doped (SLC type) cathode (bottom curve, circles) and 300-ns laser pulse.

The high charge that can be generated with the new gradient-doped strained-layer GaAsP-GaAs cathodes [3] is illustrated in Fig. 4. Note the absence of saturation effects in the upper curve, which is generated using a 100-ns laser pulse. Due to the longer pulse length of 300-ns, the current density in the lower curve is much less than the upper one for the same laser energy. For both curves, the laser illuminates a 20-mm diameter spot on the cathode. The parameters achieved for the currently running E158 using this new cathode are compared with those required for NLC in Table 3.

**Table 3**. Comparison of Parameter Achieved in E158 with NLC Requirements

| Source Parameter | E158 | NLC |
| --- | --- | --- |
| Electrons per macropulse | $8 \times 10^{11}$ e$^-$ | $27 \times 10^{11}$ e$^-$ |
| Repetition rate | 120 Hz | 120 Hz |
| Macropulse length | 270 ns | 266 ns |
| Micropulse spacing | DC | 1.4 ns |
| Polarization | ~80% | ≥80% |
| Intensity jitter | <0.5% | 0.5% |
| Transverse jitter | 5% of spot size | 22% $\sigma_x$, 50% $\sigma_y$ |

# 4 CONCLUSIONS

The SLAC polarized electron source has proven to be an outstanding performer since its initial commissioning in 1992. The source produces a charge for the presently running E158 that is close to that required for an NLC macropulse. Even higher charge has been demonstrated. Work continues at SLAC to develop a reliable low-temperature activation technique to provide more flexibility for the gradient-doped cathodes. Higher polarization is being explored, principally using superlattice semiconductor structures.